\documentclass[reprint, amsmath,amssymb, aps, prl]{revtex4-2}
\usepackage[utf8]{inputenc}
\usepackage{graphicx}
\usepackage{bm}
\usepackage{xcolor}
\usepackage[normalem]{ulem}
\usepackage[ruled,lined]{algorithm2e}
\usepackage{setspace}
\usepackage{comment}
\usepackage{adjustbox}

\usepackage{url}

\usepackage{breakurl}
\usepackage[breaklinks]{hyperref}

\definecolor{crimson}{HTML}{DC143C}
\definecolor{perfect_green}{HTML}{4FBF26}

\begin{document}

\title{Mathematical modeling of urban sprawl}

\author{Marc Barthelemy$^{1,2,*}$}
\affiliation{$^1$Université Paris-Saclay, CNRS, CEA, Institut de Physique Théorique, 91191 Gif-sur-Yvette, France}
\affiliation{$^2$Centre d'Analyse et de Mathématique Sociales (CNRS/EHESS), 54 Avenue de Raspail, 75006 Paris, France}
\affiliation{$^*$Corresponding author \texttt{marc.barthelemy@ipht.fr}}

\author{Ulysse Marquis$^{3,4}$}
\affiliation{$^3$Fondazione Bruno Kessler, Via Sommarive 18, 38123 Povo (TN), Italy}
\affiliation{$^4$Department of Mathematics, University of Trento, Via Sommarive 14, 38123 Povo (TN), Italy}

\date{\today}

\begin{abstract}

Urban land cover doubled between 1985 and 2015, yet the spatial dynamics of urban form remain under-quantified, despite its importance for sustainability, infrastructure planning, and climate risk. Urban expansion is a non-equilibrium process shaped by interactions between population growth, infrastructure, institutions, and market failures—rendering static and equilibrium models inadequate. We review key challenges and modeling approaches, focusing on partial differential equation (PDE) frameworks. Borrowed from statistical physics, PDEs capture spatial heterogeneity, anisotropy, stochasticity, and feedbacks between land use and transport networks. Integrating economic and institutional factors remains a major challenge for policy relevance. We propose a research agenda that bridges remote sensing, urban economics, and complexity science to develop dynamic, empirically grounded models of urban expansion.

\end{abstract}

\maketitle

\section*{1. Introduction}

\subsection*{1.1. Research question}

Urban sprawl refers to the spatial expansion of cities into peripheral areas. It occurs in most cities worldwide and is associated with numerous, predominantly negative consequences (see below). Figure~\ref{fig:1} illustrates the case of London, showing that urban expansion proceeds at varying speeds across different time periods and spatial directions, leading to diverse urban shapes. The challenge is to quantitatively characterize this growth, identify its driving forces, describe the evolving urban form, and ideally derive an equation that captures these spatio-temporal dynamics. The core question is therefore how to model the growth of the urban surface within a rigorous mathematical framework.
\begin{figure}
\centering
\includegraphics[width=0.49\textwidth]{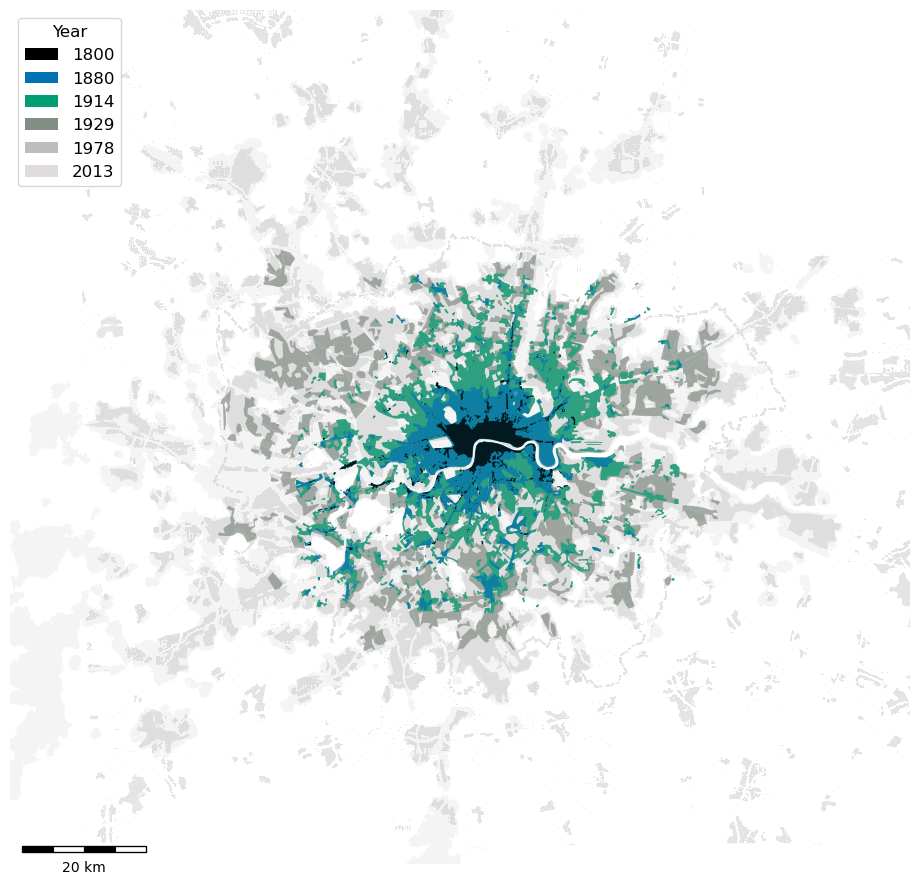}
\caption{Illustration of urban sprawl in London from 1800 to 2013. Data from \cite{angel2012atlas,angel2016atlas}; see also \cite{AtlasUrbanExpansionHistorical}.}
\label{fig:1}
\end{figure}

In this perspective, we argue that partial differential equations (PDEs)—a natural and well-established tool in physics for describing surface growth—are particularly well suited for this task. PDEs have been successfully used to model a wide range of growth phenomena, including bacterial colonies \cite{santalla2018nonuniversality}, tumors \cite{bru1998super,bru2003universal,huergo2012growth}, and turbulent liquid crystals \cite{takeuchi2010,Takeuchi_2012}, and are characterized by scaling laws and universality classes \cite{barabasi1995fractal, odor2004}. Yet, despite their generality, they have not been systematically applied to the study of urban sprawl.

More broadly, this modeling framework is not limited to urban systems and can be adapted to a wide range of complex systems. These systems are composed of many interacting components whose collective behavior gives rise to emergent patterns that are not easily reducible to individual actions. Such systems—including cities, ecosystems, or financial markets—typically exhibit nonlinear dynamics, multiscale interactions, and adaptive feedback processes. In the urban context, complexity science seeks to understand how large-scale spatial structures and socio-economic patterns emerge from decentralized, bottom-up decisions made by individuals, firms, and institutions.

Within this broader context, PDEs offer a powerful bridge between complexity science, urban economics, and remote sensing by capturing the spatio-temporal evolution of densities, flows, and potentials. In urban economics, PDEs can model how agents relocate or develop land based on local utilities, commuting costs, or housing preferences—translating micro-level decisions into macro-scale spatial patterns. From a complexity perspective, such models reveal how emergent phenomena—such as monocentric versus polycentric structures or the onset of urban sprawl—arise through feedback loops and diffusion-like mechanisms. PDEs also facilitate integration with remote sensing data, which provide high-resolution spatial information on built-up areas, transportation networks, and land-use changes. These models can be solved—at least numerically—to predict the shape of the growing urban surface or to reproduce known urban stylized facts, such as Clark’s exponential density decay. As we will see in this perspective, they indeed offer a very flexible framework able to integrate various ingredients in a clear transparent mathematical form. More precisely, to characterize this growth and shape quantitatively, we introduce a local density field $\rho(x,t)$, which may represent population density, built-up surface density, or other relevant urban quantities at location $x$ (typically a two-dimensional spatial vector). The variable $t$ generally denotes time, but it could also represent another quantity that captures the system’s evolution, such as total population or GDP. The central question we aim to address is then: how does $\rho(x,t)$ evolve over time and space, and what mechanisms govern its dynamics? A natural approach is to model this evolution using a partial differential equation of the form
\begin{align}
\frac{\partial \rho(x,t)}{\partial t} = F(\rho, x, t, \dots)
\end{align}
where the function $F$ encapsulates the underlying processes driving urban expansion. This function may depend on the density $\rho(x,t)$, space $x$, time $t$, and possibly other variables or parameters; it may also include nonlinearities, stochastic components, spatial correlations, or memory effects. Identifying a suitable form for $F$ is thus essential for understanding urban growth dynamics, with economic considerations playing a central role in determining the relevant terms of the equation.

In this perspective paper, we review several candidate models and approaches that aim to specify the function $F$. Once such a model is formulated, the corresponding equation can be solved—analytically or numerically—and its predictions compared with empirical data. A validated model not only provides insight into the dominant mechanisms driving urban growth but also allows for the quantification of their respective influences, spatial and temporal scales, and interactions. Ultimately, such a framework can serve as a powerful tool for assessing the long-term impacts of urban planning and policy decisions in a systematic and quantitative manner.

\subsection*{1.2. Key issues}

Understanding urban sprawl requires addressing several foundational challenges, both conceptual and empirical. These range from how a `city' is defined to how we represent its spatial and temporal evolution, and how we extract and interpret empirical regularities (stylized facts).\\

\begin{figure*}
\centering
\includegraphics[width=0.9\textwidth]{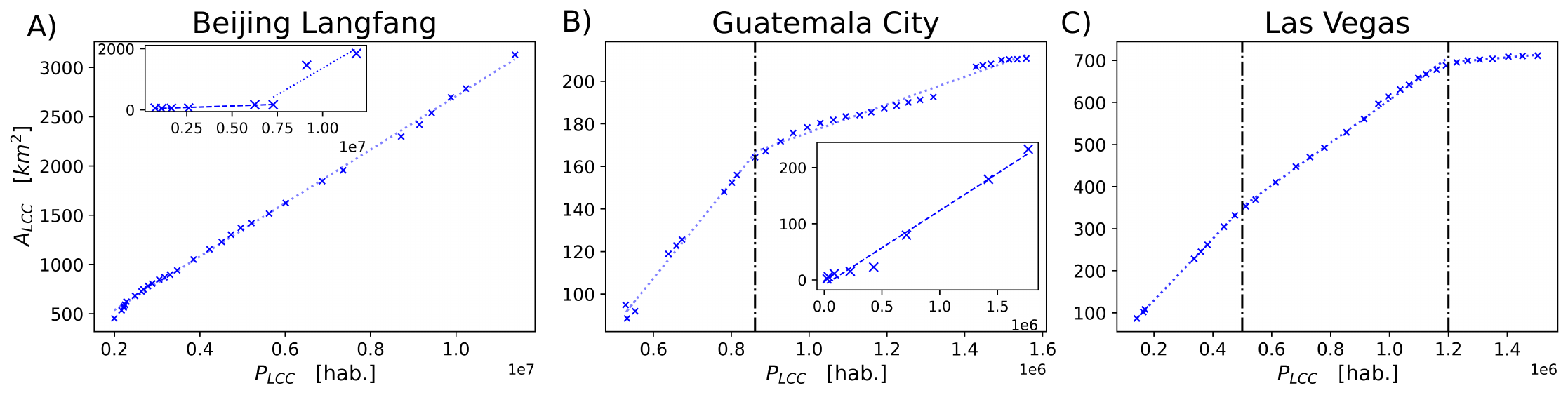}
\caption{Growth patterns of the largest connected urban area as a function of population (1985–2015). (A) Constant-density scaling (Beijing), (B) density increase (Guatemala City), (C) saturation (Las Vegas). Inset: historical density trends. Figure from \cite{marquis2025universal}.}
\label{fig:marquis1}
\end{figure*}

\paragraph{Problem 1: Defining the City}\mbox{}

A classical problem \cite{dong2024defining} is the very definition of a `city' or `urban area' that is ambiguous and varies across studies. Administrative definitions are often arbitrary and may not reflect the true extent of built-up or functionally connected regions. To overcome this, Rozenfeld et al. \cite{rozenfeld2008laws} introduced the City Clustering Algorithm (CCA), which constructs cities based on contiguous populated or built-up grid cells, bypassing arbitrary administrative boundaries. This morphological definition allows for comparative analysis across countries and is essential for identifying universal urban features. However, applying such algorithms consistently over time remains challenging due to the evolving and fragmented nature of the built environment.\\

\paragraph{Problem 2: Describing Urban Evolution}\mbox{}

Modeling the temporal evolution of urban form requires selecting appropriate variables to capture city structure. Traditional urban economics typically focused on internal variables such as the rent or wage profile, often assuming a monocentric city organized in a circular shape. However, real cities deviate significantly from this idealized model: large urban areas generally consist of a dense core surrounded by suburbs or satellite towns of varying sizes. 

Here, we focus primarily on the spatial morphology of cities, and a quantitative description generally involves three key variables: the total urbanized area $A(t)$, the evolving boundary or contour of the core built-area expressed in polar coordinates $r(\theta, t)$, and the spatial population density $\rho(x, t)$ (where $x$ is in general a two-dimensional vector). These variables allow us to track both global expansion (e.g., total area) and local heterogeneities (e.g., anisotropic growth). The increasing availability of high-resolution spatial data makes it possible to reconstruct the evolution of these features over time.

As we will see in the rest of this paper, most modeling approaches have traditionally focused on the evolution of the local population density $ \rho(x, t) $ \cite{marquis2025modeling}, while recent empirical studies \cite{marquis2025universal} have characterized urban growth through the evolution of the city's boundary $r(\theta, t)$. These variables are not independent—one may, for instance, define $r$ as the location where the density crosses a given threshold—but their mathematical treatment can differ significantly and may depend strongly on the chosen representation.\\

\paragraph{Problem 3: Time vs. population as the fundamental variable}\mbox{}

A central modeling question is whether the evolution of urban form should be viewed primarily as a function of time $t$ or of population $P$. While time naturally captures planning horizons, policy interventions, and infrastructure investments, many urban processes are endogenously driven by population dynamics. In this sense, population may represent a more fundamental variable. Models that link urban geometry to $P$ can offer greater predictive power, especially when demographic data is more reliable or more readily available than historical spatial data. Moreover, population-based models may uncover common mechanisms across cities, potentially leading to universal growth laws. If such universality exists, it is more likely to be found in the relationship between urban form and population, rather than between urban form and time. This is due in part to the fact that time typically evolves linearly—or at least continuously—whereas population does not. The relationship between $P(t)$ and $t$ can be highly irregular, shaped by exogenous factors such as migration, economic cycles, or policy shifts. As a result, population growth trajectories are often city-specific and non-universal. In contrast, the functional form of urban expansion as a function of $P$ may be similar across cities, reflecting shared mechanisms of spatial development. Still, since population itself evolves differently across cities, we may observe diverse urban trajectories even if the underlying growth law is universal. We emphasize that this framing is not exhaustive: other variables—such as GDP per capita or accessibility—may also play important roles in shaping urban growth.\\

\paragraph{Problem 4: Toward New Stylized Facts}\mbox{}

Empirical studies of urban sprawl were long constrained by limited data, with observations often fragmented in space and time. Clark’s classic model \cite{clark1951urban}—exponential decay of population density with distance—described early monocentric cities, but is less relevant for today's polycentric and anisotropic urban areas.

Recent advances in data availability have enabled the identification of new stylized facts. Remote sensing datasets like the Global Human Settlement Layer (GHSL) \cite{ghsl2020} now provide consistent global coverage of built-up areas over three decades. Historical sources such as maps and directories have also been digitized, enabling long-term reconstructions of urban form \cite{angel2017atlas} and activity \cite{gravier2024typology}. Integrating these spatial datasets with demographic information remains a challenge, but ongoing efforts promise further progress.

This convergence of data supports a more rigorous empirical foundation for modeling urban dynamics. Marquis et al. \cite{marquis2025universal} identify universal patterns such as piecewise growth at constant density (Fig.~\ref{fig:marquis1}). They also show that the average radial extent of cities scales with population as $r(\theta, P) \sim P^{\mu(\theta)}$, with $\mu(\theta)$ fluctuating around $1/2$—consistent with constant-density growth. Their work distinguishes two growth regimes: diffusion-driven expansion in low-growth contexts and coalescence of clusters in high-growth settings. Understanding these regimes and their transitions is essential for robust urban modeling.

\section*{2. Urban sprawl}

\subsection*{2.1. Definition}

Urbanization has been one of the most significant land use transformations in recent history. While urban areas still cover only a few percent of the terrestrial surface, they concentrate most of the global population, infrastructure, and economic activity. At the same time, cities are hotspots for CO$_2$ emissions, environmental degradation, and social inequalities. Between 1985 and 2015, global urban land cover increased by approximately $80\%$, with an average rate of $10{,}000\,\mathrm{km}^2$/yr—surpassing the total increase from 1970 to 2000 \cite{seto2011meta}. Urban land has grown faster than population: from 1990 to 2000, the global rate of land expansion ($3.7\%$/yr) was more than twice that of population growth ($1.6\%$/yr) \cite{angel2011dimensions}. Historical data show that most large cities have expanded much more in area than in population, with many growing more than sixteen-fold over the 20th century. This pervasive sprawl—extending cities into peripheral or rural areas—is a global, long-term phenomenon, especially pronounced in fast-developing regions such as Asia and Africa.

Projections for 2050 suggest that global urban land could triple compared to 2000, reaching $1.8$ million\,km$^2$ or more under declining density scenarios \cite{angel2011dimensions}. In developing countries, built-up area may increase four- to sixfold. Main drivers include population and income growth, land availability, and transport costs, but their importance varies: GDP dominates in high-income countries, while population growth is key in lower-income contexts \cite{seto2011meta}. Much of this expansion threatens environmentally sensitive zones, raising exposure to climate and social risks.

Urban expansion patterns are highly diverse, shaped by planning policies, geography, and historical paths. Cities can grow through continuous expansion, leapfrogging, or scattered peripheral clusters \cite{bhatta2010analysis}. A common trajectory involves an urban core expanding outward while spawning disconnected clusters that later coalesce \cite{marquis2025universal}. This variety underscores the need for comparative, quantitative approaches to describe and model urban forms. Figure~\ref{fig:1} illustrates the historical urban growth of London from 1800 to 2013, showing diffusion and coalescence processes.

Understanding the mechanisms of urban expansion is essential for addressing sprawl and promoting sustainability. Without coordinated policies, current trends are likely to persist. In this context, quantitative and modeling approaches provide essential tools to identify drivers, predict outcomes, and guide interventions \cite{marquis2025modeling}.

\subsection*{2.2. Consequences}

Urban sprawl has numerous well-documented negative impacts \cite{grimm2008global,hahs2009global}, especially in terms of environmental degradation. Rapid land consumption outpaces population growth in many regions, leading to habitat loss, reduced biodiversity, and fragmented ecosystems \cite{grimm2008global,hahs2009global}. Green spaces become isolated, undermining ecological resilience (see Table~\ref{tab:sprawl_consequences} for a recap of urban sprawl consequences).

Sprawl also alters local climates. Low-density development intensifies the urban heat island effect through the replacement of vegetation with impervious surfaces, reducing evaporative cooling and raising energy demands \cite{arnfield2003two}. Air quality worsens due to heavy reliance on private vehicles, increasing vehicle kilometers traveled (VKT) and emissions of NO$_x$, particulates, and greenhouse gases, with major consequences for health and climate \cite{bereitschaft2013urban,stone2008urban}.

Water management challenges stem from increased surface runoff and reduced groundwater recharge, as impermeable surfaces dominate the urban landscape \cite{booth1997urbanization}. Expanding stormwater infrastructure across low-density areas is costly and often degrades water quality.

Public health also suffers. Sprawl is associated with higher rates of respiratory disease, heat-related illness, and sedentary behavior due to car dependence and reduced walkability \cite{frumkin2002urban}. Mental and physical health outcomes worsen in such environments.

Economically, sprawl leads to disproportionately high infrastructure costs. Public expenditure on roads, utilities, and services must cover large, sparsely populated zones \cite{burchell2002costs,litman2023evaluating,carruthers2003urban}. Three key market failures underpin inefficient expansion \cite{brueckner2000urban}: the undervaluation of open space, unpriced congestion externalities, and developers ignoring infrastructure costs \cite{bertaud_book}. These failures, compounded by macro trends like rising incomes and cheap transport, distort land use decisions and result in unsustainable forms. In the U.S., decentralized development correlates with higher per capita public spending compared to compact urban patterns \cite{burchell2002costs,squires2002urban}.

Beyond the immediate costs, environmental damage from urban sprawl can also affect how cities grow and function over the long term. As sprawl spreads, it consumes important natural resources—such as fertile land, green spaces, and ecosystems—that help cities provide clean air, water, food, and climate regulation. This loss reduces the ability of cities to remain productive and efficient \cite{costanza1997value, dasgupta2001human}. When these natural systems are degraded, it becomes harder and more expensive to support basic needs like agriculture, water supply, or energy \cite{rockstrom2009safe}. In response, cities may try to develop new technologies to compensate, or they may end up using even more land and resources—putting further pressure on the environment and economy. If no solutions are found, cities risk needing more inputs just to maintain the same output. These self-reinforcing problems, or feedback loops, show why it's important to include environmental impacts when modeling urban growth. Doing so helps us evaluate which development paths are more sustainable in the long run \cite{bateman2014economic, dasgupta2021economics}.

\begin{table}[htbp]
\centering
\caption{Key consequences of urban sprawl, across multiple domains.}
\begin{adjustbox}{max width=\columnwidth}
\begin{tabular}{|p{2.7cm}|p{3.8cm}|p{6.5cm}|}
\hline
\textbf{Domain} & \textbf{Mechanism} & \textbf{Impact (with references)} \\
\hline
Land use \& biodiversity & Excessive land take; fragmentation of green space & Habitat loss, biodiversity decline, and disrupted ecosystems \cite{grimm2008global,hahs2009global} \\
\hline
Climate \& heat & Impervious surfaces replace vegetation & Stronger urban heat island, reduced cooling, higher energy demand \cite{arnfield2003two} \\
\hline
Air quality & Car reliance and increased VKT & More NO$_x$, PM, GHGs; worsened air quality and health risks \cite{bereitschaft2013urban,stone2008urban} \\
\hline
Water systems & Runoff increase; less infiltration & Poorer water quality, reduced groundwater recharge, higher drainage costs \cite{booth1997urbanization} \\
\hline
Public health & Car dependence, low walkability & More respiratory illness, sedentary behavior, heat-related risks \cite{frumkin2002urban} \\
\hline
Infrastructure costs & Services extended over low-density zones & High per capita costs for roads, utilities, and services \cite{burchell2002costs,litman2023evaluating,carruthers2003urban} \\
\hline
Market failures & Unpriced externalities (congestion, open space, infra) & Inefficient land use and unsustainable patterns \cite{brueckner2000urban,bertaud_book} \\
\hline
Environmental productivity & Loss of fertile land, ecosystem services, and green infrastructure & Reduced natural capital impairs long-term productivity; may increase resource use or technological dependence \cite{costanza1997value,dasgupta2001human,rockstrom2009safe,bateman2014economic,dasgupta2021economics} \\
\hline
\end{tabular}
\end{adjustbox}
\label{tab:sprawl_consequences}
\end{table}

\subsection*{2.3. Drivers of Urban Sprawl}

Urban expansion arises from interacting demographic, economic, and infrastructural drivers that vary by region and income level. In China, GDP per capita growth explains much of urban land expansion, while in India and Africa, population growth dominates \cite{seto2011meta}. High-income countries show slower expansion more tied to economic than demographic change, though regional differences persist. For instance, population growth has a stronger role in North America than in Europe, where urban containment policies are more common. Still, much of the observed variability is unexplained by GDP or population, suggesting roles for hidden drivers such as capital flows, informal economies, regulations, and transport costs \cite{seto2011meta}.

Urban sprawl also reflects land costs and individual preferences. Suburban and peri-urban areas are typically more affordable, attracting both developers and residents. Higher-income households often accept longer commutes in exchange for larger homes and green space, while developers build low-density housing with limited transit access, reinforcing car dependence.

Inadequate planning exacerbates these trends, especially where infrastructure fails to match residential growth. For example, 1990s Chicago saw fragmented suburban development disconnected from transit and employment centers \cite{mcdonald2000employment}.

Infrastructure investment in peripheral areas is another key driver of urban sprawl. Roads, utilities, and public services increase the attractiveness of outlying zones. Empirical studies show strong links between infrastructure provision and urban form. For example, a study of Canadian cities found a robust non-linear relationship between population density $\rho$ and road length per resident $L_{\mathrm{tot}}/P$, showing that low-density development requires disproportionately more road infrastructure \cite{cleveland2020shorter}. Consistently, \cite{louf2014congestion} derived the scaling relation 
$L_{\mathrm{tot}}/P\sim 1/\sqrt{\rho}$, 
in agreement with the empirical findings of \cite{cleveland2020shorter}.

Crucially, urban form and transport co-evolve: infrastructure expansion enables low-density growth, which in turn drives further transport investment. This feedback loop is well-documented \cite{wegener2004landuse}, and formalized in co-evolutionary models \cite{xie2009modeling}. A full understanding of sprawl thus requires modeling both land use dynamics and mobility networks, a topic addressed in the next section.

\section*{3. Mathematical modeling}

\subsection*{3.1. What for?}

Mathematical modeling refers to the construction of formal representations—typically in the form of equations—that describe the behavior of a system and can be solved or simulated to yield quantitative predictions. In the context of urban dynamics, such models make it possible to compute meaningful indicators such as elasticities, response functions, scaling exponents, etc. More generally, modeling provides a rigorous framework to link assumptions about mechanisms with observable outcomes. When validated against empirical data, a model demonstrates that it captures the essence of the phenomenon: it identifies the dominant mechanisms driving system evolution and reproduces the main trends, while secondary processes appear as noise \cite{batty1974urban,batty2013new}.

It is important to recognize that cities are inherently non-equilibrium systems. They evolve under the influence of endogenous processes—such as investment, migration, and innovation—and are subject to exogenous shocks, including policy interventions or environmental disruptions \cite{pumain2018evolutionary, barthelemy2016urban}. While static or equilibrium-based approaches can be useful in capturing certain stylized facts or dominant patterns, they may be limited in their ability to describe transitions, critical thresholds, or path-dependent dynamics. In this context, spatially explicit dynamic models, formulated for example as systems of PDEs, offer a complementary and more detailed analytical framework. They are particularly well suited to analyze how spatial structures emerge, evolve, and respond to perturbations in settings far from equilibrium such as in surface growth or ecology \cite{barabasi1995fractal,shigesada1997biological}.

Crucially, modeling provides both predictive and explanatory power. It enables the study of how local interactions—such as neighborhood-level land-use decisions or infrastructure investments—scale up to shape global urban forms \cite{batty2013new}. By incorporating spatial diffusion, network effects, and delayed feedbacks, models can reproduce complex spatio-temporal patterns observed in real cities \cite{makse1998modeling, west2017scale}. In this way, mathematical modeling serves not only as a theoretical framework but also as a practical tool for hypothesis testing, policy analysis, and scenario exploration.

In summary, mathematical modeling is an essential component of urban science. It offers a means to move beyond description toward mechanistic understanding and predictive insight, especially in systems as complex, adaptive, and far from equilibrium as cities.

\subsection*{3.2. Many approaches}

Several modeling paradigms have been developed to study urban sprawl, differing in abstraction, spatial resolution, and treatment of human behavior and infrastructure. These include agent-based models (ABMs), spatial interaction models, cellular automata, digital twins, and continuous approaches such as partial differential equations (PDEs) \cite{marquis2025modeling}.

ABMs simulate location choices of individual agents based on preferences, costs, and accessibility \cite{wegener2004overview}. While they capture heterogeneity and emergent patterns, they remain computationally demanding and hard to scale. Spatial interaction models, such as gravity models \cite{wilson1971family}, link spatial structure to flows but lack detailed spatial and behavioral dynamics. Cellular automata describe urban growth through local rules on grids \cite{batty1997cellular}, offering spatial insight but often omitting economic and infrastructural feedbacks.

Digital twins integrate data and simulations to create real-time virtual representations of cities, yet require stronger theoretical foundations for effective application \cite{caldarelli2023role}. Finally, continuous models using PDEs provide a powerful framework to capture diffusion, anisotropy, and feedbacks in urban systems. Their analytical tractability and ability to model spatial heterogeneity make them well-suited to study the evolution of urban form, and they are the main focus of this paper.

\subsection*{3.3. Urban economics: The Alonso-Muth-Mills model}

The Alonso--Muth--Mills (AMM) model is the foundational framework in urban economics for describing the spatial structure of monocentric cities. Developed independently by Alonso \cite{alonso1964location}, Muth \cite{muth1969cities}, and Mills \cite{mills1967aggregative}, it explains how land use and population density vary with distance from a central business district (CBD), based on a trade-off between accessibility and land consumption. It was later extended to dynamic settings \cite{wheaton1982urban}, notably in Wheaton’s model of urban growth. Given its influence, we briefly summarize the AMM framework and highlight its limitations for understanding urban sprawl.

\subsubsection*{3.3.1. Core model and results}

The AMM model assumes a monocentric city with all employment in the CBD. Identical households maximize utility by choosing a residential location, trading off land and composite consumption against commuting cost $\tau r$. Land is rented competitively, and households allocate income $Y$ between land $h$ and consumption $c$:
\begin{align}
\max_{c, h} \; u(c, h) \quad \text{s.t.} \quad Y = c + h(r) + \tau r.
\end{align}
Under Cobb--Douglas preferences $u(c,h) = c^{1-\beta} h^\beta$, the equilibrium density $\rho(r) = 1/h(r)$ decreases with distance as
\begin{align}
\rho(r) \propto (Y - \tau r)^{\frac{1 - \beta}{\beta}}.
\end{align}
When $\tau r \ll Y$, this simplifies to an exponential decay:
\begin{align}
\rho(r) \approx \rho_0 \exp(-\alpha r), \quad \text{with } \alpha = \frac{(1 - \beta)\tau}{\beta Y}.
\end{align}
The urban fringe is determined where urban rent equals agricultural rent, $h(R) = C_A$.

This model explains the observed negative gradients in density and rent. It predicts compact urban forms when commuting is costly, and sprawl when transport is inexpensive. However, it overlooks several real-world features—such as polycentricity, path dependence, housing stock persistence, and non-equilibrium dynamics—that are crucial for understanding urban sprawl.

\subsubsection*{3.3.2. Dynamic extension and limitations}

Wheaton \cite{wheaton1982urban} extended the Alonso–Muth–Mills model to a dynamic setting by introducing a framework of urban growth under perfect foresight. Assuming durable residential capital, development becomes an intertemporal allocation problem: land is developed in the period that maximizes the discounted value of future rents, accounting for evolving construction costs, income, transport costs, and population growth.

This model yields several results. Land prices always decline with distance from the center, but density may not. The direction of development can be either inside-out or outside-in, depending on historical trajectories. Rising incomes or falling transport costs produce conventional inside-out growth, while adverse trends may lead to leapfrogging or peripheral development. Simulations show that density gradients steepen with population growth and flatten when incomes fall, emphasizing that spatial patterns depend on path-dependent dynamics rather than static equilibrium \cite{wheaton1982urban}.

Despite these advances, dynamic AMM models such as Wheaton's retain fundamental flaws. One of the core issues lies in the treatment of land and development. In the original AMM formulation, every site within the city is used at its optimal intensity: there are no vacant parcels, no obsolete or underused structures, and no scope for redevelopment. The model lacks a development industry, and there are no landowners making profit-based decisions about when to demolish and rebuild. Land is continuously and instantaneously reallocated, as though directed by a benevolent social planner \cite{murray2017throw}. This assumption becomes especially problematic in comparative static applications, where any marginal change---such as the arrival of a new resident, an increase in construction efficiency, or a shift in agricultural land value---implies that the entire city is instantly rebuilt from scratch. All buildings are simultaneously demolished and replaced with a new optimal configuration, which is clearly at odds with how real cities evolve. In reality, urban change is incremental, the housing stock is persistent, and adjustments occur gradually rather than instantaneously. In this context, the AMM framework---even when extended dynamically with perfect foresight---struggles to capture the irregularities, frictions, and historical contingencies that shape real urban growth. Its equilibrium assumptions and monocentric spatial structure make it ill-suited for analyzing complex, decentralized, and path-dependent urban dynamics.

To move beyond these limitations, alternative modeling approaches are needed. In particular, partial differential equation (PDE) models provide a promising framework, dispensing with the assumption of global equilibrium and allowing for continuous, decentralized, and history-dependent urban evolution. PDEs can represent the spatial dynamics of density, land value, and accessibility fields, incorporating anisotropies, non-local effects, and coupling with infrastructure. Unlike AMM models, PDEs do not require a single city center or a representative household, offering a more flexible and realistic basis for studying the evolution of urban structure.

\section*{4. Partial differential equations approaches to urban sprawl}

\subsection*{4.1. Background: The Physics of Surface Growth and Universality Classes}

\begin{figure}
\centering
\includegraphics[width=0.45\textwidth]{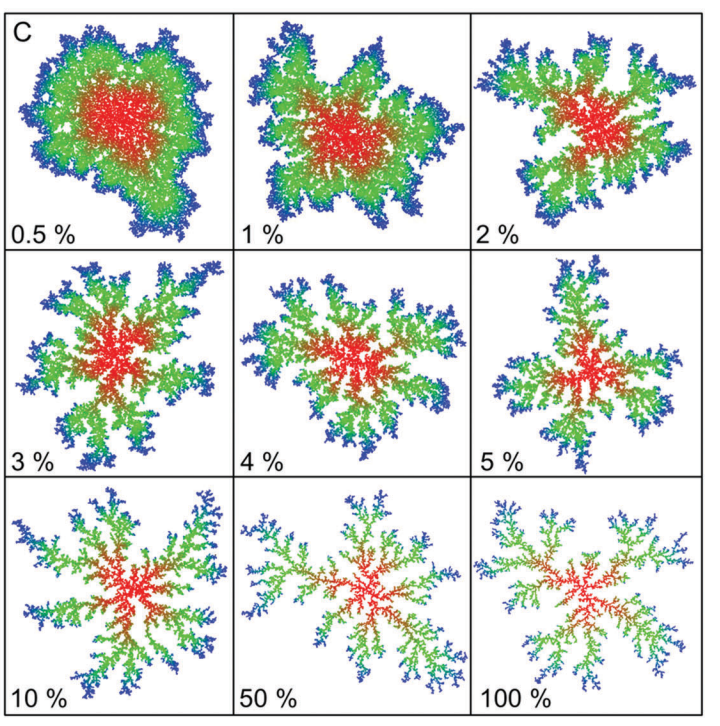}
\caption{Two-dimensional diffusion-limited aggregates with varying sticking probabilities. As adhesion increases, patterns become more branched. From \cite{heath2018visualization}.}
\label{fig:sticky}
\end{figure}

Surface growth \cite{barabasi1995fractal} is a major topic in non-equilibrium statistical physics, with applications ranging from materials to biological systems. It focuses on how interfaces evolve over time due to diffusion, aggregation, noise, and curvature. Simple local rules can generate complex global patterns (Fig.~\ref{fig:sticky}).
Most studies considered the growth along one direction and described as a surface whose central object is the height function $h(x, t)$ (where $x$ denotes the position and can be a d-dimensional vector). The evolution of this quantity is described by stochastic PDEs, and two canonical models are:

\paragraph{(i) Edwards–Wilkinson (EW) Equation \cite{edwards1982surface}}
\begin{equation}
\frac{\partial h}{\partial t} = \nu \nabla^2 h + \eta(x, t),
\end{equation}
with surface tension $\nu$ and Gaussian noise $\eta$ modeling thermal fluctuations.

\paragraph{(ii) Kardar–Parisi–Zhang (KPZ) Equation \cite{kardar1986dynamic}}
\begin{equation}
\frac{\partial h}{\partial t} = \nu \nabla^2 h + \frac{\lambda}{2} (\nabla h)^2 + \eta(x, t).
\end{equation}
Here, the nonlinearity $(\nabla h)^2$ accounts for slope-dependent growth, leading to richer scaling behavior.

Both models predict scaling laws for the surface roughness $w(L,t)$, the standard deviation of $h$ over size $L$:
\begin{align}
w(L,t) &\sim t^\beta \quad \text{for } t \ll L^z,\\
w(L,t) &\sim L^\alpha \quad \text{for } t \gg L^z,
\label{eq:alpha}
\end{align}
with exponents $\alpha$, $\beta$, and $z = \alpha/\beta$ defining a universality class. In particular, the 
exponent $\alpha$ characterizes the roughness of the surface by quantifying how height fluctuations scale with distance. A larger value of $\alpha$ indicates a rougher or more irregular surface, while smaller values correspond to smoother interfaces. These concepts extend to radial growth by replacing $h$ with the frontier position $r(\theta, t)$ \cite{marquis2025universal}.\\

\paragraph{Universality and Interdisciplinary Reach}\mbox{}

\begin{figure}
\centering
\includegraphics[width=0.45\textwidth]{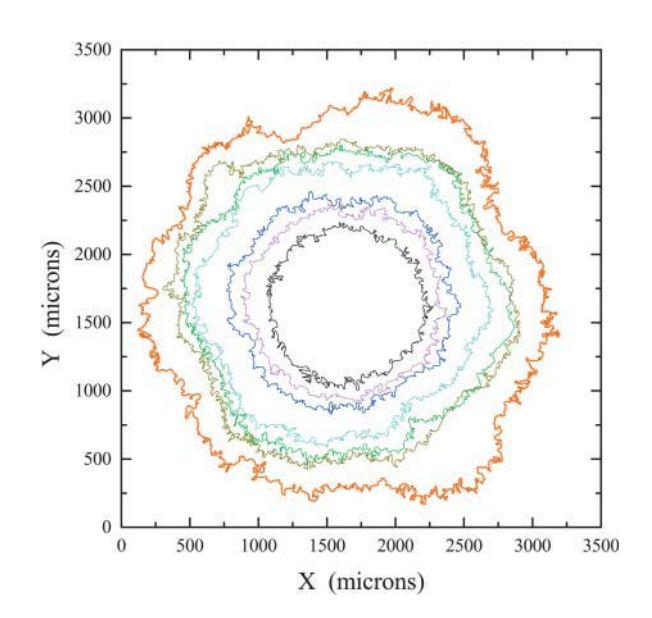}
\caption{Growth contours of a cell colony. The evolving tumor boundary exhibits scale-invariant morphology. Adapted from \cite{bru2003universal}.}
\label{fig:bru2003}
\end{figure}

Universality implies that systems with different microscopic rules—bacterial colonies \cite{wakita1997self}, tumors \cite{bru2003universal}, crystal surfaces—can share the same scaling exponents. The KPZ class appears in contexts from flame fronts to turbulent interfaces. This robustness allows classification of growth types using just a few dominant mechanisms. In biology, growth patterns—like bacterial colonies or tumors—can be modeled with KPZ-like or diffusion-limited aggregation models (DLA). Figure~\ref{fig:bru2003} shows tumor contours with roughening behavior consistent with surface growth theory.\\

\paragraph{Toward a Physics of Urban Growth}\mbox{}

Urban growth resembles biological front propagation: cities expand due to endogenous (e.g. population pressure) and exogenous (e.g. infrastructure) forces. The urban-rural frontier may follow dynamics akin to KPZ or EW growth, exhibiting roughening, anisotropy, or memory. Identifying the relevant universality class could simplify urban modeling by reducing complexity to a few exponents. Recent empirical studies provide evidence of scaling laws in urban form. In particular, a local roughness exponent—similar to the exponent $\alpha$ defined in Eq.~\ref{eq:alpha}—$\alpha_{\text{loc}} = 0.54 \pm 0.03$ appears robust across cities \cite{marquis2025universal}, suggesting a form of universality in small-scale urban boundary fluctuations. This analogy opens new directions for urban modeling, using PDEs to describe spatial density or boundary evolution. Such models bypass equilibrium assumptions and capture complex, history-dependent urban dynamics. By linking statistical physics with urban studies, we gain predictive tools for understanding and planning future urban expansion.

\subsection*{4.2. A first attempt: the isolated city}

One of the first partial differential equation (PDE) approaches to modeling urban population density was proposed by Ishikawa in~\cite{ishikawa1980new}. As in many early attempts, his objective was to reproduce the empirical result of Clark \cite{clark1951urban}. In particular, as a city’s population grows, the population density gradient tends to flatten over time, reflecting the outward spread of urban form.

In the model proposed in \cite{ishikawa1980new}, birth and death processes of population are assumed absent, and the change of residence is mainly due to two factors. First, a centripetal movement towards the city center which is due to the demand for convenience and is described by a potential that has its minimum at the city center (consistent for example with the increase of transport costs). In this view, without such a potential, no cities could ever be formed. Second, an isotropic movement from the central part of a city by a diffusion process describing the demand for an avoidance of residential crowding. The diffusion coefficient is thus supposed to be an increasing function of population density and is called in the paper `crowding pressure effect' which corresponds to the population pressure effect in ecology. 

In mathematical terms, the population density flux
$J(r,t)$ is then given by
\begin{equation}
J(r,t) = -\nabla(D\rho) - \rho \nabla V,
\end{equation}
The first term on the right-hand side in this equation means isotropic movement due to a diffusion process (with diffusion coefficient $D$). The minus sign means that the movement is directed to low population density areas. The second term on the right-hand side denotes centripetal movement
toward the city center. The quantity $V(r)$ is a potential per person, and $-\rho\nabla V$ is a contribution to the density flux by the centripetal force at population density $\rho$. $V(r)$ is an increasing function of distance from the center. In addition, it is assumed that the diffusion coefficient is a linear function of the population density
\begin{equation}
D(\rho) = \alpha + \beta \rho.
\end{equation}
where $\alpha$ is a positive parameter called intrinsic diffusion coefficient, and $\beta$ measures the avoidance of crowding. The diffusion coefficient thus consists of a positive constant and an additional term proportional to population density that reflects the avoidance of crowding. 

Birth and death processes being absent, the equation
of continuity (coming from the conservation of the total number of residents) reads
\begin{equation}
\frac{\partial \rho}{\partial t} + \nabla \cdot J=0.
\end{equation}
and the boundary condition is $J(r\to\infty)=0$. Combining all these equations, we obtain 
\begin{align}
    \frac{\partial \rho}{\partial t} = \nabla^2(D\rho)+\nabla(\rho\nabla V) .
\end{align}
We recognize in the first term the Laplacian that corresponds to diffusion and the second term the action of the force driven by the potential gradient.

In one dimension, the stationary solution~$\rho_s$ follows
\begin{equation}
\log \rho_s(x) + \frac{2 \beta}{\alpha} \rho_s(x) + \frac{V(x)}{\alpha} = C \,,
\end{equation}
and without crowding ($\beta = 0$), the density reads
\begin{align}
    \rho_s(x)\propto \mathrm{e}^{-V(x)/\alpha}
\end{align}
which reduces to Clark's model for a potential of the form $V(x)\propto |x|$. With crowding, the central density decreases and the behavior is not a pure exponential anymore. In the two-dimensional isotropic case (where the potential depends on the distance $r$ to the center $V=V(r)$), results are similar. 

This work was the first to propose a partial differential equation linking behavioral mechanisms—such as centripetal movement and crowding effects—to urban population patterns. It shows how individual behaviors can produce measurable macroscopic structures. However, the model depends on strong assumptions and reproduces only the exponential decay of density. The use of an unspecified potential $V(x)$ and a density-dependent diffusion limits internal consistency and reduces explanatory power.

\subsection*{4.3. Including congestion}

Bracken and Tuckwell \cite{bracken1992simple} proposed a dynamic PDE model of urban growth that incorporates four processes: local growth, diffusion into surrounding areas, congestion effects, and migration flows. The city is modeled as monocentric and isotropic around a CBD.

Local growth follows logistic dynamics:
\begin{align}
\frac{\partial\rho}{\partial t} = k\rho(\sigma - \rho),
\end{align}
while spatial spread is modeled by diffusion, yielding a Fisher-KPP-type equation \cite{fisher1937wave,kolmogoroff1988study}:
\begin{align}
\frac{\partial\rho}{\partial t} = D \Delta \rho + k\rho(\sigma - \rho).
\end{align}

Congestion reduces growth as central areas densify. Assuming radial symmetry, the population within radius $r$ is:
\begin{align}
N(r,t) = \int_0^r 2\pi r' \rho(r',t)\,dr',
\end{align}
and congestion is modeled via a nonlinear inhibition term $\sim \rho N(r)$:
\begin{equation}
\frac{\partial \rho}{\partial t} = D \nabla^2 \rho + k \rho(\sigma - \rho) 
- \beta \rho \int_0^r \rho(r',t)\,dr'.
\label{eq:bracken}
\end{equation}

Migration is imposed through boundary conditions:
\begin{equation}
\lim_{r \to 0} r \frac{\partial \rho}{\partial r} = \alpha,
\end{equation}
with initial condition $\rho(r,0) = \phi(r)$.

To simplify, the model is analyzed in 1D (e.g., along a coastline):
\begin{align}
\begin{cases}
\frac{\partial \rho}{\partial t} = D \frac{\partial^2 \rho}{\partial x^2} 
+ k\rho(\sigma - \rho) - \beta \rho \int_0^x \rho(x')\,dx'\\
\rho(x,0) = \phi(x)\\
\frac{\partial \rho}{\partial x}(0,t) = \alpha.
\end{cases}
\label{eq:brackendyn_1d}
\end{align}

In the absence of diffusion ($D=0$), the steady-state satisfies:
\begin{equation}
k(\sigma - \rho) = \beta \int_0^x \rho(x')\,dx' 
\quad \Rightarrow \quad \rho(x) = \sigma e^{-\beta x / k},
\end{equation}
predicting an exponential density decay. The total population is $P = \sigma k / \beta$, so increasing $P$ flattens the curve. In 2D with $D=0$, the profile becomes Gaussian:
\begin{align}
\rho(r) = \sigma e^{-\pi(\beta/k)r^2},
\end{align}
a form more consistent with empirical data showing reduced density at the center \cite{clark1951urban}.

\begin{figure}
\centering
\includegraphics[width=0.45\textwidth]{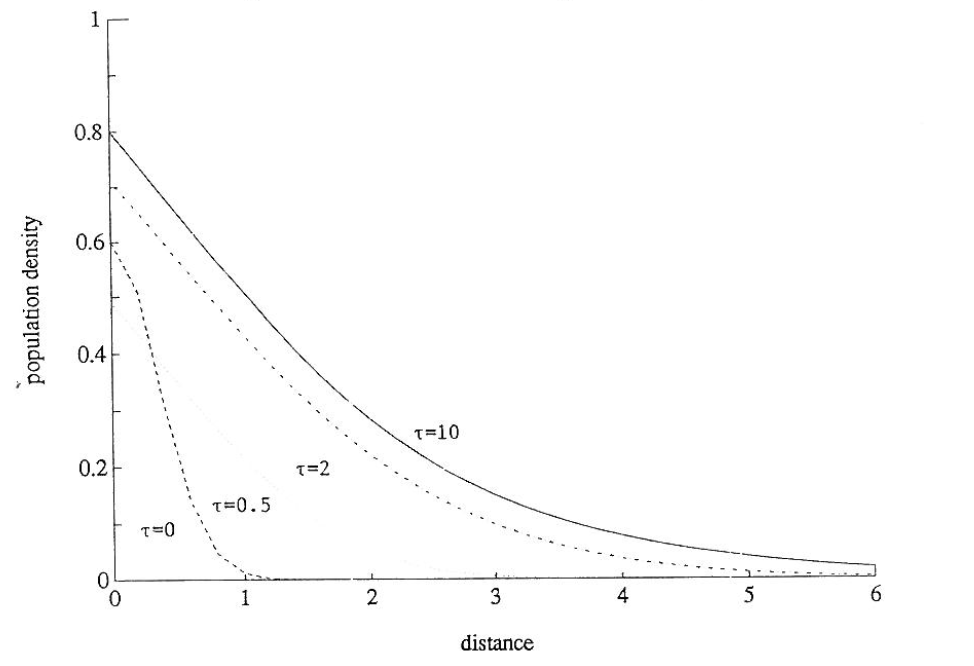}  
\caption{Time evolution of population density, showing convergence to steady state. Adapted from \cite{bracken1992simple}.}
\label{fig:bracken_time}
\end{figure}

For $D > 0$, a steady-state solution satisfies (in 1D)
\begin{align}
D \frac{d^2\rho}{dx^2} + k\rho(\sigma - \rho) - \beta\rho\int_0^x \rho(x')\,dx' = 0.
\end{align}
whose particular solution is
\begin{align}
\rho(x) = \left( \sigma + \frac{D\beta^2}{k^3} \right) e^{-(\beta/k)x},
\end{align}
but general solutions require numerical integration. Numerical simulations (Fig.~\ref{fig:bracken_time}) show convergence to a steady-state profile and a declining density gradient over time, in agreement with observations \cite{clark1951urban, ishikawa1980new}.

\begin{figure}
\includegraphics[width=0.4\textwidth]{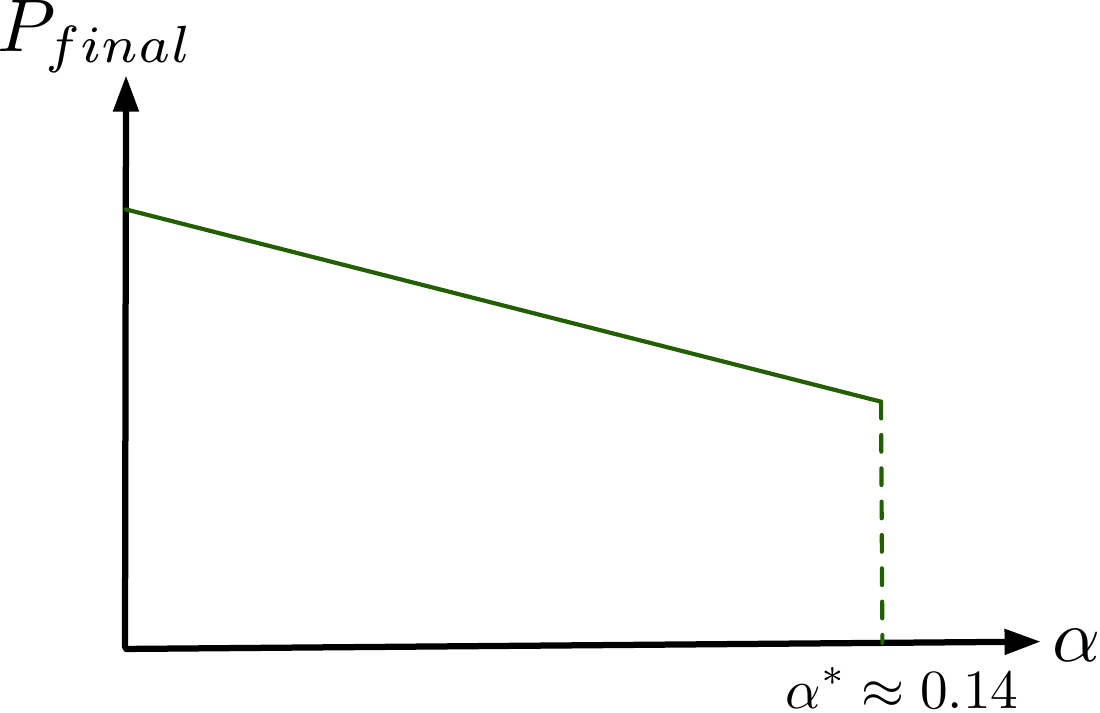}  
\caption{Final population $P_{\text{final}}$ vs emigration rate $\alpha$. Below $\alpha^* \approx 0.14$, the city persists. Above, it collapses. From \cite{bracken1992simple}.}
\label{fig:bracken}
\end{figure}

The model also explores other phenomena such as city persistence versus extinction by varying emigration rate $\alpha$. For initial condition
\begin{align}
\rho(x,0) = \sigma c \exp\left( -(\beta/k)^2 x^2 / \gamma \right),
\end{align}
numerical results show a threshold $\alpha^* \approx 0.14$: for $\alpha > \alpha^*$, the city collapses (Fig.~\ref{fig:bracken}), while $\alpha < \alpha^*$ yields survival.
Initial conditions also matter: low initial density can lead to extinction, while slightly higher values ensure survival. These results suggest a critical line in the $(c,\alpha)$ phase space.

While the model primarily aims to reproduce exponential density decay, its PDE framework with congestion and migration demonstrates how minimal ingredients can yield rich spatial dynamics. Despite limited calibration, it extends earlier approaches \cite{ishikawa1980new} and provides a flexible foundation for modeling urban evolution.

\subsection*{4.4 Including services}

Whiteley et al. \cite{whiteley2022modelling} proposed a model for urban structure formation based on the spatial dynamics of population and services. Their integro-differential framework couples population density $\rho(x,t)$ with service density $s(x,t)$, incorporating the idea that proximity to services increases attractiveness, while overly dense service areas are less desirable. They define attractiveness as
\begin{equation}
A(x,t) = (1 - s(x,t))\int K(x - y)s(y,t) dy,
\end{equation}
where $K(x)$ is a Gaussian kernel reflecting spatial influence. Individuals relocate based on attractiveness and distance, with probability
\begin{align}
P(y \to x) = \rho(y) A(x) F(d(x,y)),
\end{align}
where $F(d)$ is a decreasing function of distance. This yields the population evolution
\begin{equation} 
\frac{d\rho}{dt} = D \int [A(x) \rho(y) - A(y) \rho(x)] F(x - y)dy.
\label{eq:w1}
\end{equation}

We now have to describe the evolution of services. It is assumed that they grow logistically with carrying capacity $\sigma(P)$ dependent on the local population
\begin{equation}
\frac{ds}{dt} = (f + g s)[\sigma(P) - s]
\end{equation}
where the carrying capacity is given by
\begin{equation} 
\sigma(P) = 1 - \exp\left[-\left( \frac{P}{\lambda} \right)^\mu \right],
\label{eq:w2}
\end{equation}
where $\lambda$ sets the population scale and $\mu$ the saturation sharpness. The quantity $P(x,t)$ is a spatial average of $\rho(x,t)$ using a Gaussian kernel. Numerical analysis of Eqs.~\ref{eq:w1}–\ref{eq:w2} shows that cities emerge at characteristic spacing (e.g., 45–50 km in the UK), consistent with empirical data. 



While validation remains mostly qualitative, this framework highlights the potential of integro-differential models to explain complex urban morphology using minimal assumptions.

\section*{5. Coevolution of the population density and transport networks}

For the sake of clarity, we devote this separate section to the integration of transportation networks into a partial differential equation framework for modeling urban sprawl.

\subsection*{5.1. Impact of the transportation network}

The spatial structure of cities and transportation networks evolves through a mutual feedback process: as cities expand, infrastructure extends to meet mobility needs, while improved accessibility shapes land markets, residential patterns, and investment decisions \cite{levinson2012network,batty2013new,barthelemy2016urban,bertaud2002note}. This co-evolution has become central to debates on sustainability, congestion, and spatial equity \cite{wegener2004landuse, verbavatz2019critical}, yet many models still treat transport and urban form in isolation, missing key interactions and path dependencies \cite{xie2009modeling}. Empirical studies show that infrastructure provision directly influences urban morphology; for instance, road networks expand disproportionately in low-density contexts \cite{cleveland2020shorter}, and accessibility drives shifts from monocentric to polycentric forms \cite{louf2014congestion}. Bertaud \cite{bertaud2002note} highlights that efficient urban development depends on the coordination of transport and land-use planning. While transit-oriented development offers a promising direction, adapting legacy urban structures remains a significant constraint. A rigorous, integrated modeling framework is thus needed to capture these feedbacks and inform long-term urban planning.

Despite this recognition, few mathematical models explicitly capture the co-evolution of transport and urban form. A more rigorous modeling framework—grounded in empirical findings and feedback mechanisms—is needed to understand spatial dynamics, anticipate long-term outcomes, and guide sustainable urban planning.

\subsection*{5.2. A simple attempt for density-network feedback}

To explain the emergence of subcenters as cities expand thourgh their road network, the authors of \cite{barthelemy2009co} proposed a generative model based on a trade-off between two competing factors: rent cost, which increases with population density, and accessibility, measured via network centrality. Urban centers (e.g., homes or businesses) are added sequentially, and a growing street network evolves to connect them efficiently. This feedback mechanism captures how both center locations and infrastructure co-evolve.

\begin{figure}
  \centering
  \includegraphics[width=0.99\linewidth]{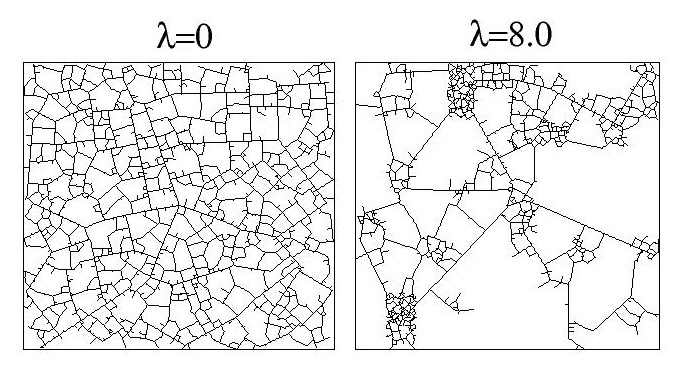}
  \caption{Networks for $\lambda=0$ (left) and $\lambda=8$ (right), with $N=500$ and $\beta=1$. Low $\lambda$ favors uniformity, as density dominates and new centers emerge in low-density areas. High $\lambda$ leads to concentration, with centrality driving growth in a few high-centrality zones. From \cite{barthelemy2009co}.}
  \label{fig:flammin1}
\end{figure}

Each city sector has area $S$ and $N(i)$ centers, yielding a density 
$\rho(i) = N(i)/S$ and rent cost 
\begin{align}
C_R(i) = A \rho(i)
\end{align}
Accessibility is captured by average betweenness centrality $g(i)$ \cite{freeman1977set} over nodes in sector $i$:
$g(i) = 1/N(i)\sum_{v \in S_i} g(v)$. The transport cost is then given by
\begin{align} 
C_T(i) = B(g_m - g(i))
\end{align}
and decreases with increasing accessibility (centrality). Assuming agents earn income $Y$, their net income in sector $i$ is
\begin{align}
K(i) = Y - C_R(i) - C_T(i).
\end{align}
The probability for a new center to locate in sector $i$ is given by
\begin{align}
P(i) = \frac{e^{\beta K(i)}}{\sum_j e^{\beta K(j)}} 
= \frac{e^{\beta A(\lambda g(i) - \rho(i))}}{\sum_j e^{\beta A(\lambda g(j) - \rho(j))}},
\end{align}
where $\lambda = B/A$ tunes the trade-off between centrality and density. According to the value of $\lambda$, two limiting behaviors emerge illustrated in Fig.~\ref{fig:flammin1}.

This model provides a minimal coupling between density and the transport network structure. The probability of placing new centers depends on both population pressure and accessibility. This approach anticipates more sophisticated models, such as the one proposed by Capel-Timms et al. \cite{capel2024angiogenic}, discussed next.

\subsection*{5.3. Coevolution of population and transport networks}

A recent contribution by Capel-Timms et al. (2024) \cite{capel2024angiogenic} introduces a compelling analogy between urban development and biological angiogenesis. The authors develop a PDE-based model in which population density evolves under economic constraints and in interaction with an adaptive transport network, capturing the coevolution of urban form and railway infrastructure. They apply this framework to the long-term development of London (1831–2011) and Sydney (1851–2011), revealing a transition from diffusion-limited densification to suburban expansion driven by increased accessibility following transport infrastructure growth. Their analysis identifies distinct phases of urban evolution—such as densification, suburbanization, decline, and regeneration—each governed by different functional forms and economic mechanisms within the model.

The partial differential equation (PDE) for population density $\rho(x, t)$, which incorporates demographic growth, spatial redistribution, and net sink/source effects, is given by
\begin{align}
\nonumber
\frac{\partial \rho(x, t)}{\partial t} = D \nabla^2 \rho(x, t) &+ r\eta(x, t)P(t)\left(1 - \frac{P(t)}{K}\right)\\
&+ R(x, t) - S(x, t),
\end{align}
Here, $D$ is the diffusion coefficient, $r$ the average growth rate, $K$ the carrying capacity, $P(t)$ the total population at time $t$, and $\eta(x, t)$ a local attractiveness factor (with $x$ a two-dimensional vector). The terms $R(x, t)$ and $S(x, t)$ represent net arrivals and departures at location $x$ and time $t$. The Laplacian term $D \nabla^2 \rho$ describes spatial diffusion from built-up areas to adjacent regions. Growth is limited by $K$ and modulated locally by $r \eta(\mathbf{x}, t)$, capturing spatiotemporal variations in urban development potential.

This spatial heterogeneity is introduced via a net income landscape as in \cite{barthelemy2009co} with $\eta(x)\propto \exp(K(x))$ where (at large times)
\begin{equation}
    K(x) = Y - C_R(x) - C_T(x),
\end{equation}
where $Y$ is the average gross income, $C_R(x)$ is the rent cost, and $C_T(x)$ is the transport cost, which depends on accessibility to the transport network. 

Living costs $C_L$ are here chosen to be
\begin{align}
    C_L(x,t)\propto\left(\int_0^t \rho(x,t)dt\right)^\tau
\end{align}
where $\tau$ is a positive parameter. The authors of \cite{capel2024angiogenic} choose the cumulative population as an indicator for the availability of buildings. 

\begin{figure*}
  \centering
  \includegraphics[width=0.99\linewidth]{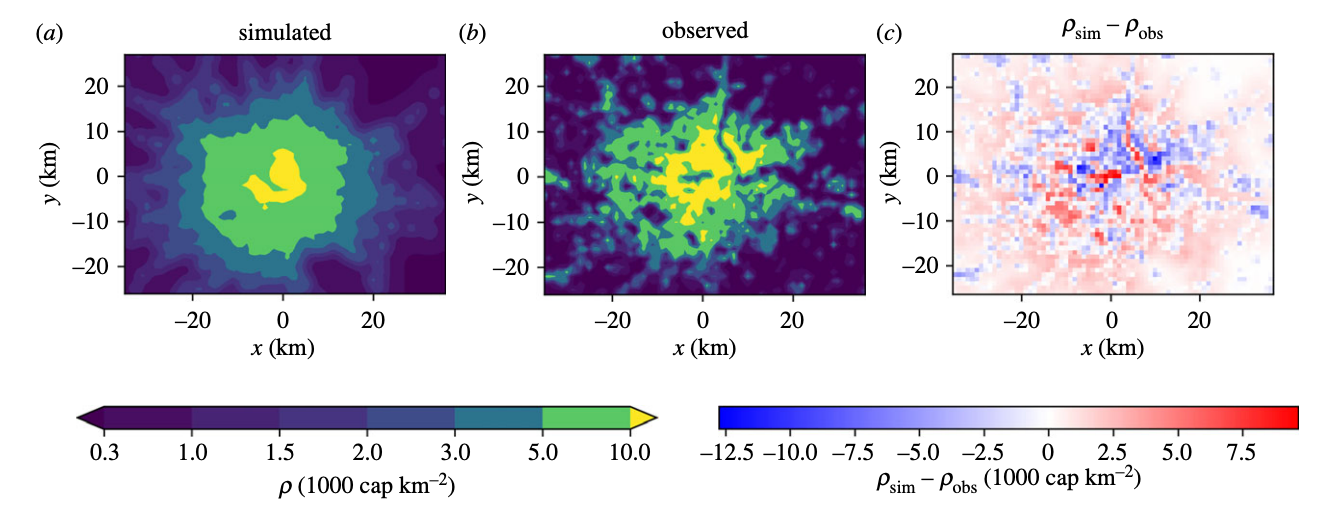}
  \caption{Simulated and observed population density in London for 2011. In (a) the simulated density, in (b) the observed density, and in (c) their difference. Figure from \cite{capel2024angiogenic}.}
  \label{fig:capel}
\end{figure*}

The transport costs $C_T(x,t)$ will depend on the current structure of the transportation network $G$. More precisely, it will be written as 
\begin{align}
    C_T(x,t)=\mathrm{const}+\mu d_{min}(x)-\nu A(x)
\end{align}
where $d_{min}(x)$ is the minimal distance from $x$ to the network, $A$ is the accessibility. There are many ways to define the accessibility, and the authors used
\begin{align}
    A_i\propto \frac{1}{\sum_jd(i,j)}
\end{align}
where $i$ is the nearest station to $x$ and $d(i,j)$ is the distance on the graph between nodes $i$ and $j$. More accessible stations have a higher value of $A$ and therefore a lower transport cost. 

Transport networks expand in response to population growth: new stations are added at location $x$ with a probability $Q(x,t)$ that increases with local population density, weighted by a power-law kernel decreasing with distance \cite{capel2024angiogenic}. These nodes are then connected based on proximity to the core. Accessibility feedback modifies future dynamics by altering $C_T$, recursively influencing $K$ and $\rho$.

The model is validated using historical data for London and Sydney. Simulations reproduce key patterns such as radial expansion, polycentricity, and commuter belt formation (see the results for the population density in London 2011 in Fig.~\ref{fig:capel}). The resulting networks display hierarchical branching and core-periphery structure, consistent with empirical findings in transport geography \cite{strano2012elementary, barthelemy2011spatial}.

This work represents an important step toward a theory that integrates the explicit dynamics of infrastructure growth with land-use feedbacks. It offers a flexible framework for scenario exploration, policy testing, and long-term forecasting of urban form and mobility infrastructure. It would also be interesting to explore simplified versions of this model that allow for analytical insights, in order to quantify the impact of infrastructure changes.

\section*{6. Discussion}

Urban sprawl is a dynamic, path-dependent process shaped by interactions between population growth, infrastructure, institutions, and market failures. Traditional equilibrium or econometric models often fail to capture these evolving feedbacks and spatial complexities.

We propose partial differential equation (PDE) models—drawn from statistical physics—as a promising framework to study urban expansion. PDEs can track local quantities like population density or built-up area over time and space, while accommodating nonlinearities, stochasticity, and multiscale interactions.

Future progress requires integrating economic and institutional factors into these models. Urban growth reflects not only physical processes but also regulatory, financial, and governance dynamics that influence infrastructure provision and land use.

Empirical validation of mathematical modeling is essential. Increasingly rich spatial datasets—such as satellite imagery—allow models to be calibrated and tested against stylized facts like scaling laws and spatial correlations. Identifying such regularities is key for building robust, generalizable theories.

Ultimately, the goal is to inform policy and planning. Understanding the co-evolution of population and infrastructure can help design more sustainable and resilient urban systems. This calls for bridging complexity science, spatial economics, and urban data analysis.

\section*{Acknowledgements}

MB thanks H. Berestycki for many discussions about this topic over the years.

\bibliographystyle{ieeetr}


\bibliography{sea_paper.bib}

@book{alonso1964location,
  title={Location and land use: Toward a general theory of land rent},
  author={Alonso, William},
  year={1964},
  publisher={Harvard university press}
}

@article{arnfield2003two,
  title={Two decades of urban climate research: a review of turbulence, exchanges of energy and water, and the urban heat island},
  author={Arnfield, A John},
  journal={International Journal of Climatology: a Journal of the Royal Meteorological Society},
  volume={23},
  number={1},
  pages={1--26},
  year={2003},
  publisher={John Wiley \& Sons, Ltd. Chichester, UK}
}

@book{angel2012atlas,
  author = {Angel, Shlomo and Parent, Jason and Civco, Daniel L. and Blei, Alexis M.},
  title = {Atlas of Urban Expansion},
  year = {2012},
  publisher = {Lincoln Institute of Land Policy},
  address = {Cambridge, MA}
}

@book{angel2016atlas,
  author = {Angel, Shlomo and Blei, Alexis M. and Parent, Jason and Lamson-Hall, Patrick and Galarza, Nicolás and Civco, Daniel L. and Qian Lei, Ru and Thom, Kristen},
  title = {Atlas of Urban Expansion—2016 Edition, Volume 1: Areas and Densities},
  year = {2016},
  publisher = {New York University, UN-Habitat, and Lincoln Institute of Land Policy},
  address = {New York, Nairobi, and Cambridge, MA}
}

@article{angel2011dimensions,
  title={The dimensions of global urban expansion: Estimates and projections for all countries, 2000--2050},
  author={Angel, Shlomo and Parent, Jason and Civco, Daniel L and Blei, Alexander and Potere, David},
  journal={Progress in planning},
  volume={75},
  number={2},
  pages={53--107},
  year={2011},
  publisher={Elsevier}
}

@misc{angel2017atlas,
  title={Atlas of urban expansion. 2016 Edition. NYU Urban Expansion Program at New York University; UN-Habitat; Lincoln Institute of Land Policy},
  author={Angel, S and Lamson-Hall, P and Madrid, M and Blei, AM and Parent, J and Galarza S{\'a}nchez, N and Thom, K},
  year={2016}
}

@misc{AtlasUrbanExpansionHistorical,
  author       = {{NYU Urban Expansion Program} and {UN-Habitat} and {Lincoln Institute of Land Policy}},
  title        = {Atlas of Urban Expansion: Historical Data},
  howpublished = {\url{http://atlasofurbanexpansion.org/historical-data}},
  note         = {Accessed: 2025-08-20},
  year         = {2016}
}

@book{barabasi1995fractal,
  title={Fractal concepts in surface growth},
  author={Barab{\'a}si, A-L and Stanley, Harry Eugene},
  year={1995},
  publisher={Cambridge university press}
}

@article{barthelemy2009co,
  title={Co-evolution of density and topology in a simple model of city formation},
  author={Barthelemy, Marc and Flammini, Alessandro},
  journal={Networks and spatial economics},
  volume={9},
  number={3},
  pages={401--425},
  year={2009},
  publisher={Springer}
}

@article{barthelemy2011spatial,
  title={Spatial networks},
  author={Barthelemy, Marc},
  journal={Physics Reports},
  volume={499},
  number={1-3},
  pages={1--101},
  year={2011},
  publisher={Elsevier}
}

@article{barthelemy2016urban,
  title={The structure and dynamics of cities: Urban data analysis and theoretical modeling},
  author={Barthelemy, Marc},
  journal={Cambridge University Press},
  year={2016}
}

@incollection{bateman2014economic,
  title={Economic analysis for ecosystem service assessments},
  author={Bateman, Ian J and Mace, Georgina M and Fezzi, Carlo and Atkinson, Giles and Turner, R Kerry},
  booktitle={Valuing Ecosystem Services},
  pages={23--77},
  year={2014},
  publisher={Edward Elgar Publishing}
}

@article{batty1974urban,
  title={Urban modelling: algorithms, calibrations, predictions},
  author={Batty, Michael},
  year={1974},
  publisher={Cambridge University Press}
}

@article{batty1997cellular,
  title={Cellular automata and urban form: a primer},
  author={Batty, Michael},
  journal={Journal of the American planning association},
  volume={63},
  number={2},
  pages={266--274},
  year={1997},
  publisher={Taylor \& Francis}
}

@book{batty2013new,
  title={The new science of cities},
  author={Batty, Michael},
  year={2013},
  publisher={MIT press}
}

@article{bereitschaft2013urban,
  title={Urban form, air pollution, and CO2 emissions in large US metropolitan areas},
  author={Bereitschaft, Bradley and Debbage, Keith},
  journal={The Professional Geographer},
  volume={65},
  number={4},
  pages={612--635},
  year={2013},
  publisher={Taylor \& Francis}
}

@inproceedings{bertaud2002note,
  title={Note on transportation and urban spatial structure},
  author={Bertaud, Alain},
  booktitle={ABCDE Conference, Washington, DC, April},
  year={2002}
}

@book{bertaud_book,
    author = {Bertaud, Alain},
    title = {Order without Design: How Markets Shape Cities},
    publisher = {The MIT Press},
    year = {2018},
    month = {12},
    isbn = {9780262349215},
    doi = {10.7551/mitpress/10671.001.0001},
    url = {https://doi.org/10.7551/mitpress/10671.001.0001},
}

@book{bhatta2010analysis,
  title={Analysis of urban growth and sprawl from remote sensing data},
  author={Bhatta, Basudeb},
  year={2010},
  publisher={Springer Science \& Business Media}
}

@article{booth1997urbanization,
  title={Urbanization of aquatic systems: degradation thresholds, stormwater detection, and the limits of mitigation 1},
  author={Booth, Derek B and Jackson, C Rhett},
  journal={JAWRA Journal of the American Water Resources Association},
  volume={33},
  number={5},
  pages={1077--1090},
  year={1997},
  publisher={Wiley Online Library}
}

@article{bracken1992simple,
  title={Simple mathematical models for urban growth},
  author={Bracken, Anthony J and Tuckwell, Henry C},
  journal={Proceedings of the Royal Society of London. Series A: Mathematical and Physical Sciences},
  volume={438},
  number={1902},
  pages={171--181},
  year={1992},
  publisher={The Royal Society London}
}

@article{bru1998super,
  title={Super-rough dynamics on tumor growth},
  author={Br{\'u}, Antonio and Pastor, Juan Manuel and Fernaud, Isabel and Br{\'u}, Isabel and Melle, Sonia and Berenguer, Carolina},
  journal={Physical Review Letters},
  volume={81},
  number={18},
  pages={4008},
  year={1998},
  publisher={APS}
}

@article{bru2003universal,
  title={The universal dynamics of tumor growth},
  author={Br{\'u}, Antonio and Albertos, Sonia and Subiza, Jos{\'e} Luis and Garc{\'\i}a-Asenjo, Jos{\'e} L{\'o}pez and Br{\'u}, Isabel},
  journal={Biophysical journal},
  volume={85},
  number={5},
  pages={2948--2961},
  year={2003},
  publisher={Elsevier}
}

@article{brueckner2000urban,
  title={Urban sprawl: Diagnosis and remedies},
  author={Brueckner, Jan K},
  journal={International Regional Science Review},
  volume={23},
  number={2},
  pages={160--171},
  year={2000},
  publisher={SAGE Publications},
  doi={10.1177/016001760002300202}
}

@book{burchell2002costs,
  title={The Costs of Sprawl—2000},
  author={Burchell, Robert W and Downs, Anthony and McCann, Barbara and Mukherji, Sahan},
  year={2002},
  publisher={National Academy Press},
  address={Washington, DC},
  note={Transit Cooperative Research Program Report 74},
  url={https://nap.nationalacademies.org/catalog/10710/the-costs-of-sprawl-2000}
}

@article{caldarelli2023role,
  title={The role of complexity for digital twins of cities},
  author={Caldarelli, Guido and Arcaute, Elsa and Barthelemy, Marc and Batty, Michael and Gershenson, Carlos and Helbing, Dirk and Mancuso, Stefano and Moreno, Yamir and Ramasco, Jos{\'e} J and Rozenblat, C{\'e}line and others},
  journal={Nature Computational Science},
  volume={3},
  number={5},
  pages={374--381},
  year={2023},
  publisher={Nature Publishing Group US New York}
}

@article{capel2024angiogenic,
  title={The angiogenic growth of cities},
  author={Capel-Timms, Isabella and Levinson, David and Lahoorpoor, Bahman and Bonetti, Sara and Manoli, Gabriele},
  journal={Journal of the Royal Society Interface},
  volume={21},
  number={213},
  pages={20230657},
  year={2024},
  publisher={The Royal Society}
}

@article{carruthers2003urban,
  title={Urban sprawl and the cost of public services},
  author={Carruthers, John I and Ulfarsson, Gudmundur F},
  journal={Environment and Planning B: Planning and Design},
  volume={30},
  number={4},
  pages={503--522},
  year={2003},
  publisher={SAGE Publications Sage UK: London, England}
}

@article{clark1951urban,
  title={Urban population densities},
  author={Clark, Colin},
  journal={Journal of the Royal Statistical Society. Series A (General)},
  volume={114},
  number={4},
  pages={490--496},
  year={1951},
  publisher={JSTOR}
}

@article{cleveland2020shorter,
  title={Shorter roads go a long way: the relationship between density and road length per resident within and between cities},
  author={Cleveland, Tristan and Dec, Paul and Rainham, Daniel},
  journal={Canadian Planning and Policy/Amenagement et Politique au Canada},
  volume={2020},
  pages={71--89},
  year={2020}
}

@article{costanza1997value,
  title={The value of the world's ecosystem services and natural capital},
  author={Costanza, Robert and d'Arge, Ralph and De Groot, Rudolf and Farber, Stephen and Grasso, Monica and Hannon, Bruce and Limburg, Karin and Naeem, Shahid and O'neill, Robert V and Paruelo, Jose and others},
  journal={nature},
  volume={387},
  number={6630},
  pages={253--260},
  year={1997},
  publisher={Nature Publishing Group UK London}
}

@book{dasgupta2001human,
  title={Human well-being and the natural environment},
  author={Dasgupta, Partha},
  year={2001},
  publisher={OUP Oxford}
}

@book{dasgupta2021economics,
  title={The economics of biodiversity: the Dasgupta review.},
  author={Dasgupta, Partha},
  year={2021},
  publisher={Hm Treasury}
}

@article{dong2024defining,
  title={Defining a city—delineating urban areas using cell-phone data},
  author={Dong, Lei and Duarte, Fabio and Duranton, Gilles and Santi, Paolo and Barthelemy, Marc and Batty, Michael and Bettencourt, Lu{\'\i}s and Goodchild, Michael and Hack, Gary and Liu, Yu and others},
  journal={Nature Cities},
  volume={1},
  number={2},
  pages={117--125},
  year={2024},
  publisher={Nature Publishing Group US New York}
}

@article{edwards1982surface,
  title={The surface statistics of a granular aggregate},
  author={Edwards, Samuel Frederick and Wilkinson, DR},
  journal={Proceedings of the Royal Society of London. A. Mathematical and Physical Sciences},
  volume={381},
  number={1780},
  pages={17--31},
  year={1982},
  publisher={The Royal Society London}
}

@article{fisher1937wave,
  title={The wave of advance of advantageous genes},
  author={Fisher, Ronald Aylmer},
  journal={Annals of eugenics},
  volume={7},
  number={4},
  pages={355--369},
  year={1937},
  publisher={Wiley Online Library}
}

@article{freeman1977set,
  title={A set of measures of centrality based on betweenness},
  author={Freeman, Linton C},
  journal={Sociometry},
  pages={35--41},
  year={1977},
  publisher={JSTOR}
}

@article{frumkin2002urban,
  title={Urban sprawl and public health},
  author={Frumkin, Howard},
  journal={Public Health Reports},
  volume={117},
  number={3},
  pages={201--217},
  year={2002},
  publisher={SAGE Publications},
  doi={10.1093/phr/117.3.201}
}

@misc{ghsl2020,
  title = {{Global Human Settlement Layer (GHSL)}},
  author = {{European Commission, Joint Research Centre (JRC)}},
  year = {2020},
  note = {\url{https://ghsl.jrc.ec.europa.eu/}},
  howpublished = {Dataset accessed 2025}
}

@article{gravier2024typology,
  title={A typology of activities over a century of urban growth},
  author={Gravier, Julie and Barthelemy, Marc},
  journal={Nature Cities},
  volume={1},
  number={9},
  pages={567--575},
  year={2024},
  publisher={Nature Publishing Group US New York}
}

@article{grimm2008global,
  title={Global change and the ecology of cities},
  author={Grimm, Nancy B and Faeth, Stanley H and Golubiewski, Nancy E and Redman, Charles L and Wu, Jianguo and Bai, Xuemei and Briggs, John M},
  journal={science},
  volume={319},
  number={5864},
  pages={756--760},
  year={2008},
  publisher={American Association for the Advancement of Science}
}

@article{hahs2009global,
  title={A global synthesis of plant extinction rates in urban areas},
  author={Hahs, Amy K and McDonnell, Mark J and McCarthy, Michael A and Vesk, Peter A and Corlett, Richard T and Norton, Briony A and Clemants, Steven E and Duncan, Richard P and Thompson, Ken and Schwartz, Mark W and others},
  journal={Ecology letters},
  volume={12},
  number={11},
  pages={1165--1173},
  year={2009},
  publisher={Wiley Online Library}
}

@article{heath2018visualization,
  title={Visualization of diffusion limited antimicrobial peptide attack on supported lipid membranes},
  author={Heath, George R and Harrison, Patrick L and Strong, Peter N and Evans, Stephen D and Miller, Keith},
  journal={Soft matter},
  volume={14},
  number={29},
  pages={6146--6154},
  year={2018},
  publisher={Royal Society of Chemistry}
}

@article{huergo2012growth,
  title={Growth dynamics of cancer cell colonies and their comparison with noncancerous cells},
  author={Huergo, Mar{\'\i}a Ana Cristina and Pasquale, MA and Gonz{\'a}lez, Pedro Horacio and Bolz{\'a}n, Agust{\'\i}n Eduardo and Arvia, Alejandro Jorge},
  journal={Physical Review E—Statistical, Nonlinear, and Soft Matter Physics},
  volume={85},
  number={1},
  pages={011918},
  year={2012},
  publisher={APS}
}

@article{ishikawa1980new,
  title={A new model for the population density distribution in an isolated city},
  author={Ishikawa, Hideaki},
  journal={Geographical Analysis},
  volume={12},
  number={3},
  pages={223--235},
  year={1980},
  publisher={Wiley Online Library}
}

@article{kardar1986dynamic,
  title={Dynamic scaling of growing interfaces},
  author={Kardar, Mehran and Parisi, Giorgio and Zhang, Yi-Cheng},
  journal={Physical Review Letters},
  volume={56},
  number={9},
  pages={889},
  year={1986},
  publisher={APS}
}

@incollection{kolmogoroff1988study,
  title={Study of the diffusion equation with growth of the quantity of matter and its application to a biology problem},
  author={Kolmogoroff, A and Petrovsky, I and Piscounoff, N},
  booktitle={Dynamics of curved fronts},
  pages={105--130},
  year={1988},
  publisher={Elsevier}
}

@article{levinson2012network,
  title={Network structure and city size},
  author={Levinson, David},
  journal={PloS one},
  volume={7},
  number={1},
  pages={e29721},
  year={2012},
  publisher={Public Library of Science San Francisco, USA}
}

@techreport{litman2023evaluating,
  title={Evaluating Transportation Land Use Impacts
Considering the Impacts, Benefits and Costs of Different Land Use Development Patterns},
  author={Litman, Todd},
  year={2023},
  institution={Victoria Transport Policy Institute},
  url={https://www.vtpi.org/landuse.pdf}
}

@article{louf2014congestion,
  title={How congestion shapes cities: from mobility patterns to scaling},
  author={Louf, R{\'e}mi and Barthelemy, Marc},
  journal={Scientific reports},
  volume={4},
  number={1},
  pages={5561},
  year={2014},
  publisher={Nature Publishing Group UK London}
}

@article{mcdonald2000employment,
  title={Employment subcenters and subsequent real estate development in suburban Chicago},
  author={McDonald, John F and McMillen, Daniel P},
  journal={Journal of Urban Economics},
  volume={48},
  number={1},
  pages={135--157},
  year={2000},
  publisher={Elsevier}
}

@article{makse1998modeling,
  title = {Modeling urban growth patterns with correlated percolation},
  author = {Makse, Hern\'an A. and Andrade, Jos\'e S. and Batty, Michael and Havlin, Shlomo and Stanley, H. Eugene},
  journal = {Phys. Rev. E},
  volume = {58},
  issue = {6},
  pages = {7054--7062},
  numpages = {0},
  year = {1998},
  month = {Dec},
  publisher = {American Physical Society},
  doi = {10.1103/PhysRevE.58.7054},
  url = {https://link.aps.org/doi/10.1103/PhysRevE.58.7054}
}

@article{marquis2025universal,
  title={Universal roughness and the dynamics of urban expansion},
  author={Marquis, Ulysse and Artime, Oriol and Gallotti, Riccardo and Barthelemy, Marc},
  journal={Phys. Rev. Lett.},
  volume={135},
  pages={187403},
  year={2025}
}

@article{marquis2025modeling,
  title={Modeling the spatial growth of cities},
  author={Marquis, Ulysse and Barthelemy, Marc},
  journal={to appear},
  year={2025}
}

@article{mills1967aggregative,
  title={An aggregative model of resource allocation in a metropolitan area},
  author={Mills, Edwin S},
  journal={The American Economic Review},
  volume={57},
  number={2},
  pages={197--210},
  year={1967},
  publisher={JSTOR}
}

@online{murray2017throw,
  author       = {Murray, Cameron},
  title        = {Throw out the standard urban economics model},
  year         = {2017},
  month        = jun,
  day          = {8},
  howpublished = {\url{https://medium.com/fresheconomicthinking/throw-out-the-standard-urban-economics-model-a495023cb7dc}}
}

@article{muth1969cities,
  title={CITIES AND HOUSING; THE SPATIAL PATTERN OF URBAN RESIDENTIAL LAND USE.},
  author={Muth, Richard F},
  year={1969}
}

@article{odor2004,
   title={Universality classes in nonequilibrium lattice systems},
   volume={76},
   ISSN={1539-0756},
   url={http://dx.doi.org/10.1103/RevModPhys.76.663},
   DOI={10.1103/revmodphys.76.663},
   number={3},
   journal={Reviews of Modern Physics},
   publisher={American Physical Society (APS)},
   author={Ódor, Géza},
   year={2004},
   month=aug, pages={663–724} }

@incollection{pumain2018evolutionary,
  title={An evolutionary theory of urban systems},
  author={Pumain, Denise},
  booktitle={International and transnational perspectives on urban systems},
  pages={3--18},
  year={2018},
  publisher={Springer}
}

@article{rockstrom2009safe,
  title={A safe operating space for humanity},
  author={Rockstr{\"o}m, Johan and Steffen, Will and Noone, Kevin and Persson, {\AA}sa and Chapin III, F Stuart and Lambin, Eric F and Lenton, Timothy M and Scheffer, Marten and Folke, Carl and Schellnhuber, Hans Joachim and others},
  journal={nature},
  volume={461},
  number={7263},
  pages={472--475},
  year={2009},
  publisher={Nature Publishing Group UK London}
}

@article{rozenfeld2008laws,
  title={Laws of population growth},
  author={Rozenfeld, Hern{\'a}n D and Rybski, Diego and Andrade Jr, Jos{\'e} S and Batty, Michael and Stanley, H Eugene and Makse, Hern{\'a}n A},
  journal={Proceedings of the National Academy of Sciences},
  volume={105},
  number={48},
  pages={18702--18707},
  year={2008},
  publisher={National Academy of Sciences}
}

@article{santalla2018nonuniversality,
  title={Nonuniversality of front fluctuations for compact colonies of nonmotile bacteria},
  author={Santalla, Silvia N and Rodr{\'\i}guez-Laguna, Javier and Abad, Jos{\'e} P and Mar{\'\i}n, Irma and Espinosa, Mar{\'\i}a del Mar and Mu{\~n}oz-Garc{\'\i}a, Javier and V{\'a}zquez, Luis and Cuerno, Rodolfo},
  journal={Physical Review E},
  volume={98},
  number={1},
  pages={012407},
  year={2018},
  publisher={APS}
}

@article{seto2011meta,
  title={A meta-analysis of global urban land expansion},
  author={Seto, Karen C and Fragkias, Michail and G{\"u}neralp, Burak and Reilly, Michael K},
  journal={PloS one},
  volume={6},
  number={8},
  pages={e23777},
  year={2011},
  publisher={Public Library of Science San Francisco, USA}
}

@book{shigesada1997biological,
  title={Biological invasions: theory and practice},
  author={Shigesada, Nanako and Kawasaki, Kohkichi},
  year={1997},
  publisher={Oxford University Press, UK}
}

@book{squires2002urban,
  title={Urban Sprawl: Causes, Consequences, \& Policy Responses},
  author={Squires, Gregory D},
  year={2002},
  publisher={Urban Institute Press},
  address={Washington, DC}
}

@article{stone2008urban,
  title={Urban sprawl and air quality in large US cities},
  author={Stone Jr, Brian},
  journal={Journal of environmental management},
  volume={86},
  number={4},
  pages={688--698},
  year={2008},
  publisher={Elsevier}
}

@article{strano2012elementary,
  title={Elementary processes governing the evolution of road networks},
  author={Strano, Emanuele and Nicosia, Vincenzo and Latora, Vito and Porta, Sergio and Barthelemy, Marc},
  journal={Scientific Reports},
  volume={2},
  pages={296},
  year={2012},
  publisher={Nature Publishing Group}
}

@article{takeuchi2010,
  title = {Universal Fluctuations of Growing Interfaces: Evidence in Turbulent Liquid Crystals},
  author = {Takeuchi, Kazumasa A. and Sano, Masaki},
  journal = {Phys. Rev. Lett.},
  volume = {104},
  issue = {23},
  pages = {230601},
  numpages = {4},
  year = {2010},
  month = {Jun},
  publisher = {American Physical Society},
  doi = {10.1103/PhysRevLett.104.230601},
  url = {https://link.aps.org/doi/10.1103/PhysRevLett.104.230601}
}

@article{Takeuchi_2012,
   title={Evidence for Geometry-Dependent Universal Fluctuations of the Kardar-Parisi-Zhang Interfaces in Liquid-Crystal Turbulence},
   volume={147},
   ISSN={1572-9613},
   url={http://dx.doi.org/10.1007/s10955-012-0503-0},
   DOI={10.1007/s10955-012-0503-0},
   number={5},
   journal={Journal of Statistical Physics},
   publisher={Springer Science and Business Media LLC},
   author={Takeuchi, Kazumasa A. and Sano, Masaki},
   year={2012},
   month=may, pages={853–890} }

@article{verbavatz2019critical,
  title={Critical factors for mitigating car traffic in cities},
  author={Verbavatz, Vincent and Barthelemy, Marc},
  journal={PLoS one},
  volume={14},
  number={7},
  pages={e0219559},
  year={2019},
  publisher={Public Library of Science San Francisco, CA USA}
}

@article{wakita1997self,
  title={Self-affinity for the growing interface of bacterial colonies},
  author={Wakita, Jun--ichi and Itoh, Hiroto and Matsuyama, Tohey and Matsushita, Mitsugu},
  journal={Journal of the Physical Society of Japan},
  volume={66},
  number={1},
  pages={67--72},
  year={1997},
  publisher={The Physical Society of Japan}
}

@book{wegener2004landuse,
  title={Land-use transport interaction: State of the art},
  author={Wegener, Michael and Fürst, Franz},
  year={2004},
  publisher={Institut für Raumplanung, Universität Dortmund}
}

@incollection{wegener2004overview,
  title={Overview of land use transport models},
  author={Wegener, Michael},
  booktitle={Handbook of transport geography and spatial systems},
  pages={127--146},
  year={2004},
  publisher={Emerald Group Publishing Limited}
}

@book{west2017scale,
author = {West, Geoffrey},
title = {Scale: The Universal Laws of Growth, Innovation, Sustainability, and the Pace of Life in Organisms, Cities, Economies, and Companies},
year = {2017},
isbn = {1594205582},
publisher = {Penguin Group , The}
}

@article{wheaton1982urban,
  title={Urban residential growth under perfect foresight},
  author={Wheaton, William C},
  journal={Journal of urban Economics},
  volume={12},
  number={1},
  pages={1--21},
  year={1982},
  publisher={Elsevier}
}

@article{whiteley2022modelling,
  title={Modelling the emergence of cities and urban patterning using coupled integro-differential equations},
  author={Whiteley, Timothy D and Avitabile, Daniele and Siebers, Peer-Olaf and Robinson, Darren and Owen, Markus R},
  journal={Journal of the Royal Society Interface},
  volume={19},
  number={190},
  pages={20220176},
  year={2022},
  publisher={The Royal Society}
}

@article{wilson1971family,
  title={A family of spatial interaction models, and associated developments},
  author={Wilson, Alan Geoffrey},
  journal={Environment and Planning A},
  volume={3},
  number={1},
  pages={1--32},
  year={1971},
  publisher={Sage Publications Sage UK: London, England}
}

@article{xie2009modeling,
  title={Modeling the growth of transportation networks: A comprehensive review},
  author={Xie, Feng and Levinson, David},
  journal={Networks and Spatial Economics},
  volume={9},
  number={3},
  pages={291--307},
  year={2009},
  publisher={Springer}
}

\end{document}